\documentclass[prl,showpacs,superscriptaddress,twocolumn]{revtex4}

\begin{document}

\title{Ascertaining the Values of $\sigma_x$, $\sigma_y$, and $\sigma_z$ 
of a Polarization Qubit}

\author{Oliver Schulz}
\affiliation{Sektion Physik, Universit\"at M\"unchen, %
Schellingstra\ss{}e 4, 80799 M\"unchen, Germany}

\author{Ruprecht Steinh\"ubl}
\affiliation{Sektion Physik, Universit\"at M\"unchen, %
Schellingstra\ss{}e 4, 80799 M\"unchen, Germany}

\author{Markus Weber}
\affiliation{Sektion Physik, Universit\"at M\"unchen, %
Schellingstra\ss{}e 4, 80799 M\"unchen, Germany}

\author{Berthold-Georg~Englert}
\affiliation{Max-Planck-Institut f\"ur Quantenoptik, %
Hans-Kopfermann-Stra\ss{}e 1, 85748 Garching, Germany}
\affiliation{Department of Physics, National University of Singapore
2~Science~Drive~3, Singapore 117\,542, Singapore} 

\author{Christian Kurtsiefer}
\affiliation{Sektion Physik, Universit\"at M\"unchen, %
Schellingstra\ss{}e 4, 80799 M\"unchen, Germany}

\author{Harald Weinfurter}
\affiliation{Sektion Physik, Universit\"at M\"unchen, %
Schellingstra\ss{}e 4, 80799 M\"unchen, Germany}
\affiliation{Max-Planck-Institut f\"ur Quantenoptik, %
Hans-Kopfermann-Stra\ss{}e 1, 85748 Garching, Germany}

\date{30 August 2002}

\begin{abstract}%
In the 1987 spin retrodiction puzzle of Vaidman, Aharonov, and Albert
one is challenged to ascertain the values of $\sigma_x$, $\sigma_y$, and
$\sigma_z$ of a spin-$\frac{1}{2}$ particle by utilizing entanglement.
We report the experimental realization of a quantum-optical version 
in which the outcome of an intermediate polarization projection is inferred
by exploiting single-photon two-qubit quantum gates.
The experimental success probability is consistently 
above the $90.2\%$ threshold of the optimal one-qubit strategy, 
with an average success probability of $95.6\%$.
\end{abstract}


\maketitle

The 1987 paper \cite{VAA} by Vaidman, Aharonov, and Albert (VAA) answered the
question of ``How to Ascertain the Values of $\sigma_x$, $\sigma_y$, and
$\sigma_z$ of a Spin-$\frac{1}{2}$ Particle'' and so showed, in the words of
Mermin, how to perform the following trick:
Alice prepares a quantum mechanical system in a certain initial state and
gives it to Bob.
Without telling Alice his choice, Bob measures either $\sigma_x$, $\sigma_y$,
or $\sigma_z$ of a spin-$\frac{1}{2}$ particle contained in the system, and
gives the system back to Alice, who makes an additional measurement.
This enables her (still not knowing Bob's choice) correctly to announce what
Bob's result was if he measured $\sigma_x$, what it was if he measured
$\sigma_y$, and what it was if he measured $\sigma_z$ \cite{M}.

Thanks to Aharonov's popularization, this
spin-retrodiction challenge became generally known as the 
\emph{Mean King's Problem}.
It embeds the VAA puzzle into a colorful tale where 
Alice is a ship-wrecked physicist
and Bob the underling of the physicist-hating despot 
who rules the remote island on which  Alice got stranded \cite{AE}. 

For the quantum-optical realization we employ the polarization of single
photons as the VAA puzzle's degree of freedom of spin-$\frac{1}{2}$ type.
The standard for judging the experimental results is set by
the optimal strategy that Alice can follow if she
manipulates solely the single polarization qubit in question.
Photon states with horizontal and vertical polarization   
($\mathsf{h}$ or $\mathsf{v}$)
are identified with the eigenstates of $\sigma_z$, 
right and left circular polarization states ($\mathsf{r}$ or $\mathsf{l}$)
with those of $\sigma_y$, and linear polarization states under $+45^\circ$ and
$-45^\circ$ ($\mathsf{+}$ or $\mathsf{-}$) with those of  $\sigma_x$,
\begin{eqnarray}
  \label{eq:1}
&  \sigma_x=\Ket{+}\Bra{+}-\Ket{-}\Bra{-}\,,&\nonumber\\
&  \sigma_y=\Ket{r}\Bra{r}-\Ket{l}\Bra{l}\,,\quad
   \sigma_z=\Ket{h}\Bra{h}-\Ket{v}\Bra{v}\,.&
\end{eqnarray}
Imagine now that Alice just prepares the photon in a certain polarization
state
--- right circular polarization ($\mathsf{r}$), say.
She can then surely infer the correct answer if Bob measures $\sigma_y$.
At the final stage, she measures ${\sigma_x+\sigma_z}$, thereby 
finding the photon
linearly polarized either half-way between $\mathsf{h}$ and $\mathsf{+}$
or half-way between $\mathsf{v}$ and~$\mathsf{-}$.
Although she lacks perfect retrodiction if Bob measured $\sigma_x$ or
$\sigma_z$, she can guess his measurement result rather well, namely with
total betting odds of $\frac{1}{3}(2+2^{-\frac{1}{2}})=90.2\%$.
In fact, one demonstrates easily \cite{demo} that 
this is the largest likelihood for guessing right that she can
achieve by such a single-qubit strategy. 

To do better than these $90.2\%$, Alice takes to heart the advice given by the
VAA trio and entangles the photon polarization with an auxiliary qubit
(which is not revealed to Bob).
Alice's final projection of the auxiliary qubit and the photon returned by Bob 
onto an entangled basis allows perfect polarization retrodiction and so
enables her to solve the Mean King's Problem.

An essential ingredient of all variants of the VAA puzzle 
and its various generalization \cite{M,B-M,AE,EA,EKW}
is that the intermediate measurement
by Bob is an ideal von Neumann measurement that finds an eigenvalue of the
observable in question and leaves the system behind in the respective 
eigenstate.
In the present optical experiment, we use projections as equivalent
replacements of von Neumann measurements~\cite{EKW'}. 
For the photons that are successfully projected by Bob,
Alice faces the original VAA problem of
determining which projection occurred at the intermediate stage. 

\begin{figure}[!t]
\epsfig{file=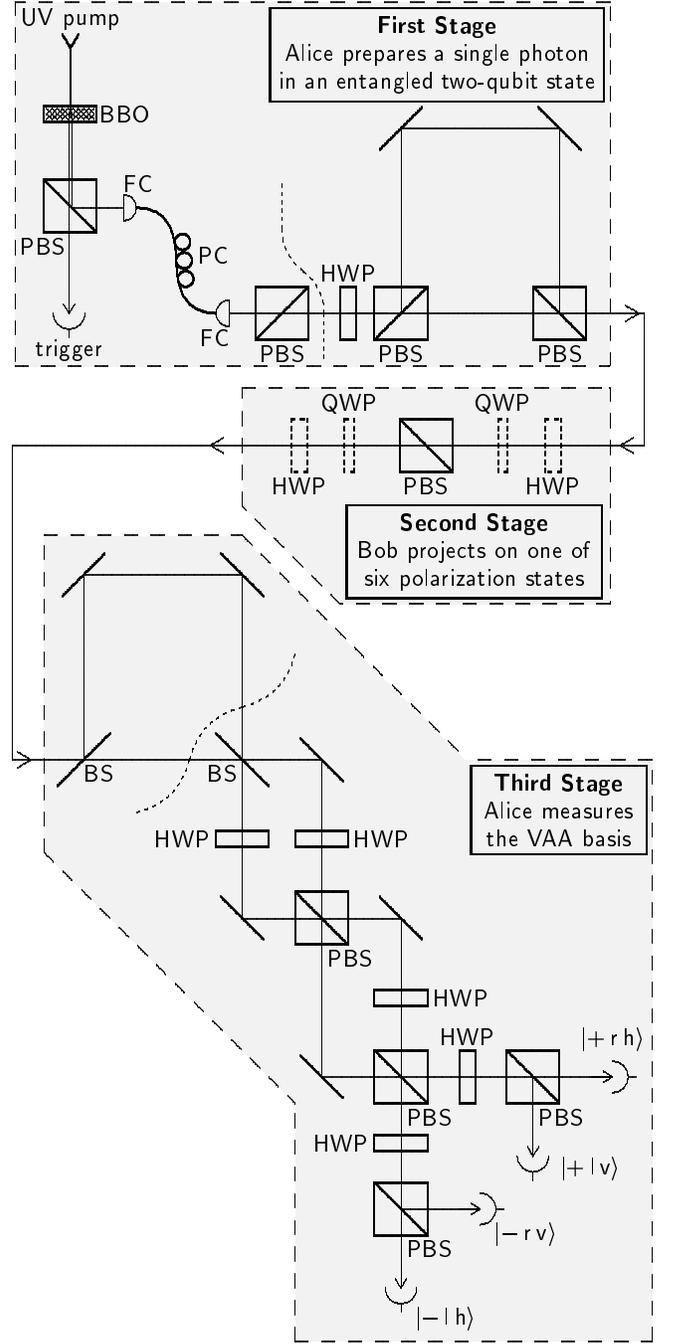,bbllx=54,bblly=233,bburx=299,bbury=743}
\caption{\label{fig:setup}%
The three stages of the quantum-optical experiment that realizes the Mean
King's Problem.}
\end{figure}

At the first stage of the experiment, 
Alice prepares the entangled two-qubit state.
In our experiment, see Fig.~\ref{fig:setup}, the auxiliary qubit
is a longitudinal spatial mode of the photon, namely the binary
alternative of being early or late ($\mathsf{E}$ or $\mathsf{L}$)
which, entangled with the polarization of the photon, is in 
the single-photon two-qubit state
\begin{equation}
  \label{eq:5}
  \Ket{init}=2^{-\frac{1}{2}}\bigl(\Ket{E,h}+\Ket{L,v}\bigr)\,,
\end{equation}
where $\Ket{E,h}$, for instance, denotes a horizontally
polarized photon that arrives early. 
This additional $\mathsf{E}$/$\mathsf{L}$ qubit is hidden from Bob who does
not know the precise instant when the photon is ready.

At the second stage,  
Bob projects the entangled state of (\ref{eq:5}) onto one of six product
states, depending on the polarization he actually selects,
\begin{equation}\label{eq:6}
\begin{array}[b]{r@{:\ \Ket{init}\to}l}
  \sigma_x&\left\{
    \begin{array}{l}
     \Ket{'+'}\equiv2^{-\frac{1}{2}}(\Ket{E,+}+\Ket{L,+})\,,\\[0.5ex]
     \Ket{'-'}\equiv2^{-\frac{1}{2}}(\Ket{E,-}-\Ket{L,-})\,,
    \end{array}\right.
\\[3ex]
  \sigma_y&\left\{
    \begin{array}{l}
    \Ket{'r'}\equiv2^{-\frac{1}{2}}(\Ket{E,r}-i\Ket{L,r})\,,\\[0.5ex]
    \Ket{'l'}\equiv2^{-\frac{1}{2}}(\Ket{E,l}+i\Ket{L,l})\,,
    \end{array}\right.
\\[3ex]
  \sigma_z&\left\{
    \begin{array}{l}
    \Ket{'h'}\equiv\Ket{E,h}\,,\\[0.5ex]
    \Ket{'v'}\equiv\Ket{L,v}\,.
    \end{array}\right.
\end{array}
\end{equation}
Then, at the third stage, 
Alice performs a measurement in which she distinguishes the
four mutually orthogonal states of the VAA basis that are given by
\begin{eqnarray}
  \label{eq:7}
\Ket{+\,r\,h}&=&2^{-\frac{1}{2}}\Ket{E,h}
+\tfrac{1}{2}i^{\frac{1}{2}}\Ket{E,v}+\tfrac{1}{2}i^{-\frac{1}{2}}\Ket{L,h}
\,,\nonumber\\
\Ket{+\,l\,v}&=&2^{-\frac{1}{2}}\Ket{L,v}
+\tfrac{1}{2}i^{-\frac{1}{2}}\Ket{E,v}+\tfrac{1}{2}i^{\frac{1}{2}}\Ket{L,h}
\,,\nonumber\\
\Ket{-\,r\,v}&=&2^{-\frac{1}{2}}\Ket{L,v}
-\tfrac{1}{2}i^{-\frac{1}{2}}\Ket{E,v}-\tfrac{1}{2}i^{\frac{1}{2}}\Ket{L,h}
\,,\nonumber\\
\Ket{-\,l\,h}&=&2^{-\frac{1}{2}}\Ket{E,h}
-\tfrac{1}{2}i^{\frac{1}{2}}\Ket{E,v}-\tfrac{1}{2}i^{-\frac{1}{2}}\Ket{L,h}
\,,
\end{eqnarray}
where $i^{\pm\frac{1}{2}}=(1\pm i)/\sqrt{2}$.
Upon detecting the second VAA state $\Ket{+\,l\,v}$, for example, 
Alice would infer that Bob projected on the $\mathsf{+}$ polarization 
if the choice was between the $\sigma_x$ alternatives of
$\mathsf{+}$ and $\mathsf{-}$, that he projected on $\mathsf{l}$ if the
choice was between $\mathsf{r}$ and $\mathsf{l}$, and on $\mathsf{v}$
if it was the $\sigma_z$ choice between $\mathsf{h}$ and $\mathsf{v}$.
Her inference is always correct because $\Ket{+\,l\,v}$
is orthogonal to $\Ket{'-'}$, $\Ket{'r'}$, and $\Ket{'h'}$ by construction.

The setup of the first stage of the experiment, in which Alice prepares the
two-qubit state (\ref{eq:5}), is sketched at the top of 
Fig.~\ref{fig:setup}. 
To generate a single photon with a well defined emission time, she first
produces a pair of simultaneously emitted photons, then detects one of them 
to record the time of emission, and uses the other for the
polarization-retrodiction experiment.
To the left of the dashed line, we have first
the nonlinear BBO crystal, in which an incoming ultraviolet (UV) 
photon is converted into a pair of co-propagating infrared photons --- one
$\mathsf{h}$ polarized, the other $\mathsf{v}$ polarized 
(parametric down conversion of type II).
This pair is split at a polarizing beam splitter (PBS), which transmits the  
$\mathsf{h}$ photon and reflects the $\mathsf{v}$ photon.
The $\mathsf{h}$ photon is detected, and this gives us the trigger 
signal by which the detectors of the third stage (bottom part of
Fig.~\ref{fig:setup}) are gated.
To ensure single-mode operation, 
the $\mathsf{v}$ photon is fed into a single-mode fiber through a fiber
coupler (FC). 
Upon emerging from the fiber through another FC the photon passes through a
PBS that selects $\mathsf{h}$ polarization.
The fiber is equipped with a polarization control (PC) to manipulate the
photon polarization such that the yield of this selection is maximized.
   
Accordingly, a photon that makes it to the dashed line at the top of 
Fig.~\ref{fig:setup} is assuredly $\mathsf{h}$ polarized. 
It passes through a half-wave plate (HWP) that changes
the polarization state to $\Ket{+}=2^{-\frac{1}{2}}(\Ket{h}+\Ket{v})$.
The photon then traverses an unbalanced Mach-Zehnder interferometer (MZI)
that has PBSs at the entry and exit ports. 
As a consequence, the $\mathsf{h}$ component takes the short way and
emerges early ($\mathsf{E}$), 
and the $\mathsf{v}$ component takes the long way and is late ($\mathsf{L}$).
The photon amplitude is thereby split longitudinally, because
the detour of about 90\,cm is longer than the coherence length 
of the photon ($\sim0.1$\,mm).
The photon is now prepared in the single-photon two-qubit state (\ref{eq:5}),
and Alice hands it over to Bob.
Without knowing the trigger time for reference, it is impossible for him to
recognize the particular preparation.
All he can see is a randomly polarized photon.

Bob has the second stage under control, 
the center part of Fig.~\ref{fig:setup}, where
he performs one of the six projections (\ref{eq:6}).
For projection on state $\Ket{'h'}$, the PBS alone suffices, since it
reflects $\mathsf{v}$ polarized photons and transmits
$\mathsf{h}$ polarized ones.
For all other projections, a suitably set HWP, or a quarter wave plate (QWP),
or both are used to turn the polarization in question into $\mathsf{h}$. 
The then transmitted photon is $\mathsf{h}$ polarized, and by passing it
through a second HWP, or QWP, or both its polarization is turned back to the
wanted one.
In this manner, each of the six projections (\ref{eq:6}) can be implemented by
Bob.
If his projection is successful, the photon is forwarded to the third stage,
otherwise Bob has to ask Alice to prepare another photon and the procedure
must be repeated~\cite{fn1}.

The third stage, the bottom part of Fig.~\ref{fig:setup}, 
begins with the conversion of
the longitudinal alternative of arriving ``early or late'' 
($\mathsf{E}$ or $\mathsf{L}$) into the transversal
alternative of moving ``downwards or to the right''
($\mathsf{D}$ or $\mathsf{R}$).
This is achieved with the aid of the beam splitter (BS) and two mirrors to the
left of the dashed curve.
The $\mathsf{E}$ component takes the detour over the mirrors (of the same
length as the one in the top part)
and becomes $\mathsf{D}$, the $\mathsf{L}$ component goes straight ahead and
becomes $\mathsf{R}$.
In the other cases --- $\mathsf{E}$ going straight ahead or $\mathsf{L}$
taking the detour --- the photon will arrive at one of the detectors
either before or after the time interval during which they are gated in
accordance with the trigger impulse from the first stage.
The path length difference of 90\,cm translates into a time delay of 3\,ns,
which can be conveniently resolved by an electronic gate window of 
1.2\,ns~\cite{fn2}. 

As soon as the $\mathsf{E}\to\mathsf{D}$, $\mathsf{L}\to\mathsf{R}$ conversion
is accomplished (and the dashed curve is reached in the bottom part
of Fig.~\ref{fig:setup}),
the VAA basis of (\ref{eq:7}) could be measured 
either with a MZI with non-polarizing BSs \cite{EKW},
or with PBSs, as was chosen here because of the easier fine tuning of
the 50:50 beam splitting.
An additional interferometer loop  
connects the conversion stage with the VAA analyzer.
The whole setup thus consists of two consecutive MZIs, where the first
has a BS at the input port and a PBS at the output port. 
The latter serves also as the input port of the second MZI, 
which has another PBS at its output port.
After emerging at one of the two output channels of the second MZI, the photon 
passes through yet another PBS and is then detected by one of four detectors.
All HWPs are oriented at $22.5^\circ$ or $-22.5^\circ$ 
such that a photon that arrives in one of
the four VAA states (\ref{eq:7}) is guided to the corresponding photodetector. 

The setup shown in Fig.~\ref{fig:setup} is schematic.
In the real implementation the mirrors of the various MZIs are retro-reflecting
prisms such that a single BS can serve as input and output component.
Additional compensator plates (not indicated in Fig.~\ref{fig:setup}) correct
for birefringence of BSs and prisms.
And each of the four interferometer loops is phase locked with a He-Ne
reference laser.

\begin{figure}[!t]
\epsfig{file=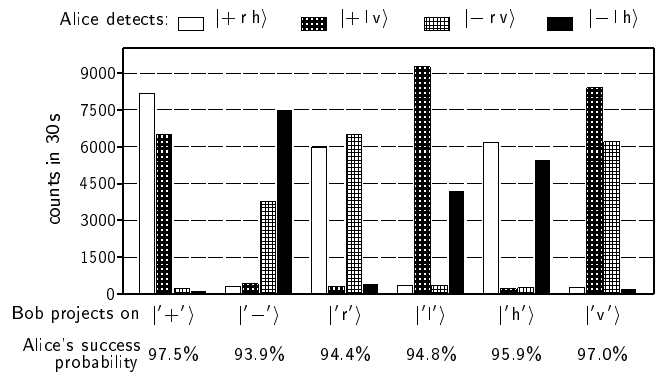,bbllx=52,bblly=600,bburx=297,bbury=742}
\caption{\label{fig:Counts}%
Outcome of a run of the experiment in which the clicks of the gated 
detectors of the third stage in Fig.~\ref{fig:setup} 
are counted for 30\,s for each of the projections (\ref{eq:6}); see text.
Alice's inferred odds for guessing Bob's projection right exceed, in each
case, the $90.2\%$ odds of the optimal single-qubit strategy,
with average odds of $95.6\pm1.2\%$.
}
\end{figure}

Once one of the detectors has fired (in the gate window), 
Alice knows immediately which
projection was performed by Bob if he chose between $\mathsf{+}$ and
$\mathsf{-}$, what it was if he chose between $\mathsf{r}$ and $\mathsf{l}$,
and  what it was if he chose between $\mathsf{h}$ and $\mathsf{v}$.
The outcome of one run of the experiment is shown in Fig.~\ref{fig:Counts}. 
After the wave plates of the second stage of Fig.~\ref{fig:setup} 
are set properly 
to effect either one of the six projections of (\ref{eq:6}),
the counts of
the gated detectors of the third stage are recorded for a duration of
30\,s.  
Upon determining the respective fractions of clicks by the wrong detectors,
one infers Alice's experimental odds for guessing Bob's projection right.
Her success probability exceeds in each case the $90.2\%$ odds of the optimal
single-qubit strategy.
On average the odds are $95.6\%$, 
with a statistical error of $\pm1.2\%$.

Imperfections of the optical elements and of their alignment result in
occasional clicks by a detector that shouldn't fire and, therefore, the
projection is not always inferred correctly, as the data reported in
Fig.~\ref{fig:Counts} show.
But we do get consistently better odds of guessing the projection 
right than the $90.2\%$ that the best single-qubit strategy would offer.
Indeed, there is a true pay-off from entangling the photon polarization
with a spatial alternative of the photon 
(first ``$\mathsf{E}$  or $\mathsf{L}$'' --- 
then ``$\mathsf{D}$  or $\mathsf{R}$''), 
and we have succeeded in realizing the Mean King's Problem by quantum-optical
means.

It may be worthwhile to state explicitly 
how the challenge would be phrased in
the particular context of our experimental setup.
First Bob would choose one of the six projections,
set the wave-plates of his stage fittingly, and
would then tell Alice to send the photon.
She follows suit, prepares a photon at the first stage,
and waits for one of the detectors of the third stage to fire.
It is important that Bob also has access to the knowledge if a photon has been
detected or not.
If no detector fires, Alice has to send a second photon, 
and a third (and fourth, fifth, \dots) if necessary.
But as soon as one photon is detected, Bob calls in Alice's guess for the
chosen projection, and this run is over.  

Rather than the VAA basis of (\ref{eq:7}), Alice can just as well measure the
other VAA basis \cite{BEKW1,BEKW3} that consists of the complementary set of
states $\Ket{-\,l\,v}$, $\Ket{-\,r\,h}$, $\Ket{+\,l\,h}$, and $\Ket{+\,r\,v}$.
In the setup of Fig.~\ref{fig:setup}, one only needs to change 
by $\frac{1}{2}\pi$ the phase of the polarizing MZI of the VAA analyzer, 
and by $\pi$ the phase of the connecting MZI loop, 
by adjusting some arm lengths.
We have performed the projection-retrodiction experiment also with 
this other VAA basis and have obtained
results similar to those reported in Fig.~\ref{fig:Counts},
with average odds of $94.7\pm1.2\%$.  

In the \emph{Mean King's Second Challenge} \cite{BEKW3}, Bob does not choose
between one of six polarization projections, but rather between one of six
unitary polarization transformations --- three pairs of two:
\begin{equation}\label{eq:6'}
\begin{array}[b]{@{\Ket{init}\to2^{-\frac{1}{2}}}l@{\Ket{init}=}l}
  (\sigma_x\pm\sigma_y)&\left\{
    \begin{array}{l}
    (\frac{1}{2}i)^{\frac{1}{2}}(\Ket{E,v}-i\Ket{L,h})\,,\\[0.5ex]
     (2i)^{-\frac{1}{2}}(\Ket{E,v}+i\Ket{L,h})\,,
    \end{array}\right.
\\[3ex]
  (\sigma_y\pm\sigma_z)  &\left\{
    \begin{array}{l}
    2^{-\frac{1}{2}}(\Ket{E,r}-i\Ket{L,l})\,,\\[0.5ex]
    -2^{-\frac{1}{2}}(\Ket{E,l}+i\Ket{L,r})\,,
    \end{array}\right.
\\[3ex]
  (\sigma_z\pm\sigma_x)&\left\{
    \begin{array}{l}
    2^{-\frac{1}{2}}(\Ket{E,+}+\Ket{L,-})\,,\\[0.5ex]
    2^{-\frac{1}{2}}(\Ket{E,-}-\Ket{L,+})\,,
    \end{array}\right.
\end{array}
\end{equation}
and Alice has to find out which of the two transformations of a pair was
actually performed after eventually being told which of the three pairs
applies.
In the corresponding experimental setup, only Bob's second stage is different,
Alice's preparation in the first stage and her detection in the third stage
remain exactly the same.
Another difference is the threshold set by the optimal single-qubit strategy.
It is only $\frac{5}{6}=83.3\%$ for the Second Challenge.
Our experimental guessing odds had average success probabilities that were
consistently in excess of this threshold, 
with average odds of $92.2\pm0.7\%$.

In summary, then, the 1987 spin-retrodiction puzzle by Vaidman, Aharonov, and
Albert --- the Mean King's Problem --- 
has been realized in the form of a quantum-optical analog, in which
one infers which polarization projection was performed on a single photon.
We have achieved success probabilities that exceed, in each channel, 
the single-qubit optimum.
Further, we have successfully implemented the Mean King's Second Challenge, in
which unitary polarization changes are performed rather than polarization
projections.
The realization of this quantum game is based on implementing single-photon
quantum logic \cite{EKW}, and is thus a first step toward more complex tasks
of all-photonic quantum computers~\cite{KLM}.

As an outlook we note that the Mean King's Problem suggests 
deterministic schemes for quantum cryptography \cite{BEKW1} and for
direct secure quantum communication ``with a publicly known key'' 
\cite{BEKW2,BEKW3}.
A demonstration experiment for the cryptography scheme is already
feasible with an apparatus that differs only by the simultaneously implemented
second VAA basis analyzer.

\begin{acknowledgments}
We thank Yakir Aharonov and Lev Vaidman for their encouragement and the
lessons learned in illuminating discussions.
This work was supported by project QuCommm (IST-1999-10033) of the 
European Union and the DFG.
\end{acknowledgments}


\begin{thebibliography}{99}

\bibitem{VAA}
L. Vaidman, Y. Aharonov, and D. Z. Albert,
\prl {\bf 58}, 1385 (1987).

\bibitem{M}
N. D. Mermin, \prl {\bf 74}, 831 (1995).

\bibitem{AE}
Y. Aharonov and B.-G. Englert,
Z. Naturforsch. \textbf{56a}, 16 (2001).

\bibitem{demo}
Alice prepares the photon in a certain polarization state specified by the
statistical operator
 $\rho=\tfrac{1}{2}(1+x\sigma_x+y\sigma_y+z\sigma_z)$
with $x^2+y^2+z^2\leq1$.
In her control measurement, she measures $\Vec{e}\cdot\Vec{\sigma}$ 
with some unit vector $\Vec{e}=(e_x,e_y,e_z)$.
After being told which intermediate measurement was performed by Bob
she bets on the outcome that contributes most to the probability that
the result actually found appears in her control measurement.
Summed over all possible scenarios,
her total odds for guessing right are  
\begin{displaymath}
\textstyle
  \frac{1}{2}+\frac{1}{6}\sum_{\zeta=x,y,z}
\max\bigl\{|\zeta|,|e_\zeta|\bigr\}
\leq\frac{2}{3}+\frac{1}{6}\sqrt{2}=90.2\%\,.
\end{displaymath}
The maximum of $90.2\%$ is
achieved for $(x,y,z)=(0,1,0)$ and 
$\Vec{e}=(2^{-\frac{1}{2}},0,2^{-\frac{1}{2}})$, 
for example, which corresponds to the situation described in the text.

\bibitem{B-M}
S. Ben-Menahem, \pra {\bf 39}, 1621 (1989).

\bibitem{EA}
B.-G. Englert and Y. Aharonov,
\pl \textbf{A284}, 1 (2001).

\bibitem{EKW}
B.-G. Englert, C. Kurtsiefer, and H. Weinfurter,
\pra \textbf{63}, 032303 (2001).

\bibitem{EKW'}
In the proposal of \cite{EKW} 
the von Neumann measurement is mimicked by following the destructive 
photodetection with a suitable unitary polarization change on another photon. 
An experiment along these lines would require very fast switches and, 
therefore, we modified the scheme and replaced measurements by projections.
See also \cite{fn2}.

\bibitem{fn1}
Bob's projection is successful in 50\% of the cases.
With perfect detectors, he would note that his preparation failed upon
registering the photon reflected from his PBS.

\bibitem{fn2}
Only 50\% of the incoming photons are detected during the gate time.
This can be avoided by replacing the first BS of the third stage by a
polarization insensitive, quickly switchable mirror.
While such a device is conceptually possible, one would lose efficiency and
signal-to-noise ratio with present-day technology.

\bibitem{BEKW3}
A. Beige, B.-G. Englert, C. Kurtsiefer, and H. Weinfurter,
``Communicating with qubit pairs''
in \textit{Mathematics of Quantum Computation}, edited by R. Brylinski and
G. Chen (CRC, Boca Raton, 2002), pp.~359--401.

\bibitem{BEKW1}
A. Beige, B.-G. Englert, C. Kurtsiefer, and H. Weinfurter,
J. Phys.\ A \textbf{35}, L407 (2002).

\bibitem{KLM}
E. Knill, R. Laflamme, and G. J. Milburn,
\nat \textbf{409}, 46 (2001).

\bibitem{BEKW2}
A. Beige, B.-G. Englert, C. Kurtsiefer, and H. Weinfurter,
Acta Phys.\ Pol.\ A \textbf{101}, 357 (2002); \textbf{101}, 901 (2002) .
   
\end{thebibliography}
\end{document}